\documentclass[runningheads]{llncs}

\newcommand{\ExtendedVersion}[1]{#1} \newcommand{\ConferenceVersion}[1]{}

\usepackage{graphicx}
\usepackage[table,xcdraw]{xcolor}
\usepackage{wrapfig}

\usepackage[makeroom]{cancel}
\usepackage{cite}

\usepackage{enumerate}

\usepackage{subcaption}
\usepackage{graphicx}

\usepackage{amsmath}
\usepackage{mymacros}
\usepackage{amssymb}
\usepackage[colorinlistoftodos,textsize=tiny]{todonotes}

\usepackage[ruled,vlined,linesnumbered]{algorithm2e}%
\usepackage{amsmath,algpseudocode}

\makeatletter
\renewcommand{\Function}[2]{%
  \csname ALG@cmd@\ALG@L @Function\endcsname{#1}{#2}%
  \def\jayden@currentfunction{#1}%
}
\newcommand{\funclabel}[1]{%
  \@bsphack
  \protected@write\@auxout{}{%
    \string\newlabel{#1}{{\jayden@currentfunction}{\thepage}}%
  }%
  \@esphack
}
\makeatother
\algdef{SE}{Begin}{End}{\textbf{begin}}{\textbf{end}}

\usepackage[colorlinks=true,%
            linkcolor=black,%
            citecolor=black,%
            urlcolor=black]{hyperref}
\usepackage[section]{placeins}
\usepackage{verbatim}
\usepackage{mathabx}
\usepackage{multirow}
\usepackage{capt-of}
\usepackage{tikz}
\usepackage{times}
\usepackage{xcolor}

\usepackage{nameref}
\newcounter{mylabelcounter}

\makeatletter
\newcommand{\labelText}[2]{%
#1\refstepcounter{mylabelcounter}%
\immediate\write\@auxout{%
  \string\newlabel{#2}{{1}{\thepage}{{\unexpanded{#1}}}{mylabelcounter.\number\value{mylabelcounter}}{}}%
}%
}

\algnewcommand\algorithmicforeach{\textbf{for each}}
\algdef{S}[FOR]{ForEach}[1]{\algorithmicforeach\ #1\algorithmicdo}
\def\univs{\mathbb{U}}
\newcommand{\univ}[1]{\univs_{\mathit{#1}}}

\SetKwInput{KwData}{Input}
\SetKwInput{KwResult}{Output}

\usepackage{xcolor}
\SetCommentSty{mycommfont}

\hypersetup{bookmarksopen=true}

\begin{document}

\mainmatter              

\title{Transforming Object-Centric Event Logs \\ to Temporal Event Knowledge Graphs}
\ExtendedVersion{
	\subtitle{(Extended Version)}
}

\titlerunning{Transforming OCEL to Temporal Event Knowledge Graphs}

\author{Shahrzad Khayatbashi\inst{1} 
 \and Olaf Hartig \inst{1} \and Amin Jalali\inst{2}}

\institute{Linköping University, Linköping, Sweden,\\
\email{(shahrzad.khayatbashi | olaf.hartig)@liu.se}
\and
Stockholm University, Stockholm, Sweden,\\
\email{aj@dsv.su.se}
}
\maketitle

\vspace{-0.5\baselineskip}
\begin{abstract}   
Event logs play a fundamental role in enabling data-driven business process analysis. Traditionally, these logs track events related to a single object, known as the case, limiting the scope of analysis. Recent advancements, such as Object-Centric Event Logs~(OCEL) and Event Knowledge Graphs~(EKG),  capture better how events relate to multiple objects. However, attributes of objects can change over time, which was not initially considered in OCEL or EKG. While OCEL 2.0 has addressed some of these limitations, there remains a research gap concerning how attribute changes should be accommodated in EKG and how OCEL 2.0 logs can be transformed into EKG. This paper fills this gap by introducing Temporal Event Knowledge Graphs~(tEKG) and defining an algorithm to convert an OCEL~2.0 log to a tEKG.

\keywords {Event Knowledge Graphs, Object-Centric Event Data, Object-Centric Process Mining}
\end{abstract}

\section{Introduction} \label{sec:intro}

\vspace{-0.5\baselineskip}

Business processes involving participants, resources, and systems can be analyzed from different perspectives~\cite{jalali2013multi}. These perspectives include different objects based on which a process can be analyzed for further improvement. Traditional analysis focuses on a single object~(a.k.a. case), making it challenging to answer questions considering multiple objects and perspectives simultaneously. Object-Centric Process Mining~(OCPM) addresses this limitation, aiming to uncover insights by capturing interrelations between objects and events in event logs. Data that includes the relation between events to multiple objects is known as Object-Centric Event Data~(OCED)~\cite{van2023object}, promising the discovery of more insights and addressing the limitations of traditional analysis methods.

Around 2020, two data models were introduced to record OCED: Object-Centric Event Log~(OCEL)\cite{ghahfarokhi2021ocel} and Event Knowledge Graph~(EKG)\cite{fahland2022process}. OCEL~1.0~\cite{ghahfarokhi2021ocel} proposed a conceptual model for event logs, enabling the recording of events related to multiple objects and facilitating the development of OCPM algorithms, e.g.~\cite{van2020discovering,gherissi2023object,jalali2022object}.
EKG presented an alternative technique to record event logs in a Knowledge Graph~\cite{esser2019storing,esser2021multi}.

Transforming logs between these formats not only enables the application of techniques developed for each format but also facilitates comparing limitations, which can be used to extend these models for further analysis. 
For instance, a recent study on transforming EKG to OCEL 1.0 highlights the lack of support in capturing relations between objects in OCEL~\cite{khayatbashi2023transforming}, a concern now addressed by OCEL 2.0~\cite{berti2023ocelspecification}.  

OCEL~2.0 extends its predecessor with support to record information on object relationships, to qualify relationships, and to capture the change of values for attributes of objects over time~\cite{berti2023ocelspecification}. This extension allows capturing the temporal value of objects in practice. As an example, the price of an item in an online webshop can change over time. If these price changes aren't accurately tracked, it becomes difficult to analyze why an item suddenly becomes popular. This is because we lack the correct price data for when customers viewed the item at different times.

While EKG has also undergone improvements, it lacks support for such temporal aspects. 
Additionally, there exists a gap in transforming OCEL~2.0 to EKG, hindering a direct comparison and tool reuse between these two formats. To fill these gaps, this paper focuses on the following research questions:
\begin{description}
	\itemsep1mm
	\item[~ \normalfont RQ1:] How can temporal aspects of object attributes be captured in an EKG?
	\item[~ \normalfont RQ2:] How can an OCEL 2.0 log be transformed into a temporal EKG?
\end{description}

To address \text{RQ1} we extend the EKG model into a model of Temporal Event Knowledge Graphs~(tEKGs). 
To address \text{RQ2} we define an algorithm for transforming an OCEL~2.0 file into our proposed tEKG representation%
. We have implemented this algorithm and provide the source code of this implementation publicly.\footnote{\url{https://github.com/shahrzadkhayatbashi/BPM2024}}

Structure of the paper:
Section~\ref{sec:background} gives a background using a running example, and Section~\ref{sec:approach} introduces tEKGs informally.
Section~\ref{sec:formalization} defines tEKG formally,
Section~\ref{sec:transformation} defines the transformation algorithm%
, and Section~\ref{sec:conclusion} concludes the paper. 

\section{Background}\label{sec:background}
\vspace{-0.5\baselineskip}

This section introduces the relevant background on EKGs based on a running example.

The example revolves around a fictional education-related process where a student failed to pass a course and must retake it the following year. In the first year, the student read instructions for an assignment and submitted it accordingly. Subsequently, the teacher decided to increase the points allocated for the assignment from 2 to 3, as it was discovered that the assignment was considerably more challenging than anticipated.
Here, we provide a high-level overview of this process to convey the essential concepts necessary for understanding EKG and tEKG. To ensure simplicity, we do not include representations of students, teachers, or other entities typically involved in such a process. A key aspect, however, is that the number of points of the assignment can change over time and must be correctly recorded for the different years.

\begin{figure}[t]
    \vspace{-0.5\baselineskip}
	\centering 
	\includegraphics[width=1\textwidth]{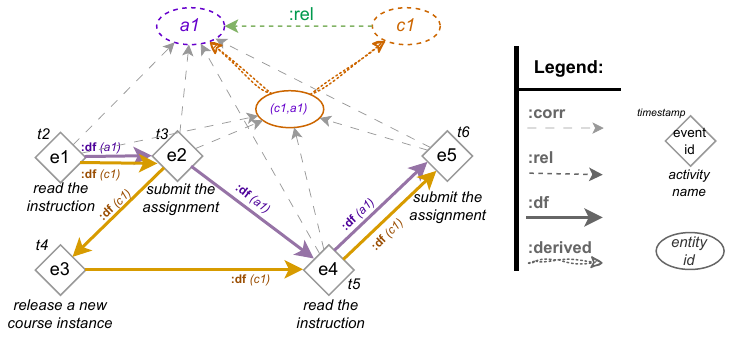}
    \vspace{-1.5\baselineskip}
        \caption{A part of an Event Knowledge Graph}    
	\label{fig:ekg}
    \vspace{-1\baselineskip}
\end{figure}

\figurename~\ref{fig:ekg} illustrates a part of an EKG using nodes and edges to record data of our running example. In EKGs, nodes can be labeled as $\mathsf{Log}$, $\mathsf{Entity}$, $\mathsf{Class}$, or $\mathsf{Event}$.  These nodes capture information about log files, objects, event types, and events, respectively.  
The label $\mathsf{Entity}$ is used for nodes representing objects in EKGs. In this paper, we use the terms "object" and "entity" interchangeably when referring to an object in OCEL and EKG, respectively.
In the figure, only nodes labeled $\mathsf{Entity}$ and $\mathsf{Event}$ are displayed. For instance, $\mathsf{c1}$ and $\mathsf{a1}$, depicted as ovals, represent the course and the assignment, respectively, in our example. An event, denoted by $\mathsf{e1}$ and shown as a diamond, captures the event of the student reading the instruction at time $\mathsf{t2}$. 
Each node can be annotated with key-value pairs called properties. For example, an assignment may have a specific number of points that students can receive upon submitting the assignment.

Edges in EKGs establish relationships among nodes, and these edges can be labeled to indicate the type of relationship between nodes. For example, relationships between entities are represented using edges labeled as $\mathsf{rel}$. In our example, $\mathsf{c1}$ is connected to $\mathsf{a1}$ via such an edge, indicating that the course has an assignment. Edges between nodes labeled $\mathsf{Event}$ and nodes labeled $\mathsf{Entity}$ are labeled $\mathsf{corr}$, and the label $\mathsf{df}$ is used for edges representing directly-follows relationships among nodes representing events.

Such directly-follows relationships
	between events
are fundamental in process mining. In EKGs, two events, say $\mathsf{e1}$ and $\mathsf{e2}$, can be connected by a $\mathsf{df}$ edge if
i)~there is a shared entity to which both events have a $\mathsf{corr}$ edge,
ii)~$\mathsf{e1}$ occurred before $\mathsf{e2}$, and
iii)~there are no other events that fulfill the first condition and occurred between $\mathsf{e1}$ and $\mathsf{e2}$.
Such a $\mathsf{df}$ edge is associated with two properties called $\mathsf{EntityID}$ and $\mathsf{EntityType}$, representing the identifier and type of the corresponding entity to which the two connected events are related via a $\mathsf{corr}$ edge. As an example, in the figure, $\mathsf{e1}$ is linked to $\mathsf{e2}$ by an edge labeled $\mathsf{df}$, with $\mathsf{a1}$ as the value of $\mathsf{EntityType}$.

In this graph, all events have a $\mathsf{corr}$ edge to $\mathsf{c1}$ since they are events that occurred within this course. However, we have not depicted these edges in this figure to avoid overwhelming complexity. Instead, we have visualized the resulting $\mathsf{df}$ edges. It is apparent that the event flow related to the assignment differs from that of the course, primarily because releasing a new course instance is unrelated to the assignment.

There are scenarios where it becomes necessary to link an object to the relationship between objects, which cannot be achieved directly in EKGs because an edge cannot connect a node to another edge. In EKGs this limitation is addressed by adding helper nodes known as reified entities.
	An example of a reified entity in \figurename~\ref{fig:ekg} is $\mathsf{(c1,a1)}$ which reifies the $\mathsf{rel}$ edge from $\mathsf{c1}$ to $\mathsf{a1}$. These two entities are connected to the reified entity by edges with the label $\mathsf{derived}$. All events connected to the aforementioned entities will also be connected to the reified entity. For more detailed information on these concepts, and on EKGs in general, we refer readers to Fahland's work~\cite{fahland2022process}.

	We emphasize that the EKG model lacks the capability to capture changes in the values of attributes of entities. In our example, the value of the $\mathsf{Points}$ property of $\mathsf{a1}$ was modified to 3 in the second year when the student retook the course. Consequently, the recorded information for the first year, where the assignment had 2 $\mathsf{Points}$, would be lost due to the overwrite. This discrepancy can lead to erroneous analysis results. The tEKG model that we propose in this paper addresses this limitation.

\section{Temporal Event Knowledge Graphs}\label{sec:approach}

\begin{figure}[b]
     \centering
     \begin{subfigure}[b]{0.47\textwidth}
         \centering
         \includegraphics[width=\textwidth]{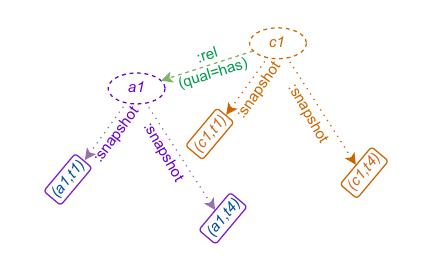}
         \caption{Creating snapshots for entities}
     \end{subfigure}
     \hfill
     \begin{subfigure}[b]{0.47\textwidth}
         \centering
         \includegraphics[width=\textwidth]{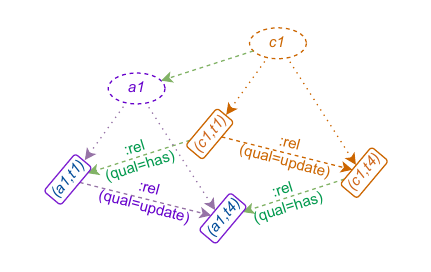}
         \caption{relating snapshots}
     \end{subfigure}
     
        \caption{Creating snapshots to capture changes in object's attributes over time}
        \label{fig:tekg:snapshots}
    \vspace{-1\baselineskip}
\end{figure}

This section introduces our proposal informally and discusses our design choices.

Our initial design choice is to ensure backward compatibility with EKG. This choice aims to facilitate the reuse of existing solutions, such as inferring missing entity identifiers~\cite{swevels2023inferring} or aggregating event knowledge graphs for task analysis~\cite{klijn2022aggregating}. Therefore, we define tEKG as an extension of the EKG model that supports all EKG features as well as additional features for handling temporal entities.

The values of attributes of entities can change over time, and there are various methods to track these changes in information systems. One approach involves recording transactions for attribute modifications, while another entails capturing snapshots of the state of an object at different points in time.
The latter method is commonly employed in data warehousing to store facts in periodic snapshot fact tables~\cite{kimball2011data}, which prioritize query performance for data analysis over transactional performance. Therefore, we have chosen a similar design choice to enhance query performance, which involves generating snapshots of entities when the values of their attributes change.

tEKGs contain multiple nodes per entity; one such node represents the entity in general, independent of the temporal dimension (i.e., exactly as in an EKG), and the other nodes represent snapshots of the entity at specific times. The identifier of each snapshot node is the combination of a timestamp and the identifier of the corresponding entity.
Figure~\ref{fig:tekg:snapshots}(a) illustrates such snapshot nodes; in particular, the course $\mathsf{c1}$ and assignment $\mathsf{a1}$ each have two snapshots at times $\mathsf{t1}$ and $\mathsf{t4}$, respectively. The relationship between each entity and its snapshots is established through edges with label~$\mathsf{snapshot}$.

\begin{figure}[!t]
	\centering 
	\includegraphics[width=0.9\textwidth]{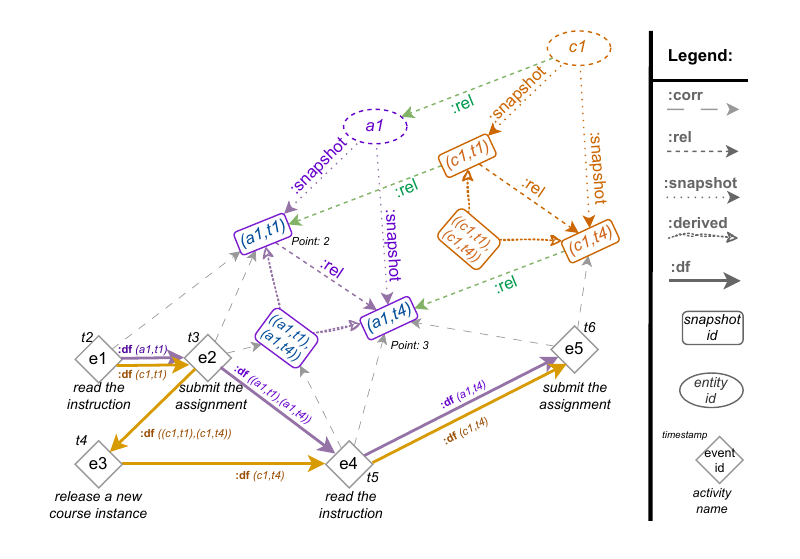}
        \caption{An example of a Temporal Event Knowledge Graph}
	\label{fig:tekg}
\vspace{-1\baselineskip}
\end{figure}

Edges with the label~$\mathsf{rel}$ can be used to capture relationships between snapshots, as illustrated in Figure~\ref{fig:tekg:snapshots}(b).
	Such an edge has an attribute named $\mathsf{qual}$ with a value of $\mathsf{update}$ if the connected snapshots are for the same entity.
 Essentially, such edges document the lifecycle of an entity within a tEKG. 
For instance, assignment~$\mathsf{a1}$ is initially created at time~$\mathsf{t1}$ and subsequently updated at time~$\mathsf{t4}$, as captured by the snapshot nodes $\mathsf{(a1,t1)}$ and $\mathsf{(a1,t4)}$, respectively.
 The edges labeled $\mathsf{rel}$ between entities will be copied to their snapshots, with the condition that each snapshot is connected only to snapshots that have existed in its lifetime.

We adapt the design choices made for EKGs to reify entities for snapshots, to connect events to snapshots, and to create directly-follows edges between events corresponding to the same snapshots. This results in additional edges in our graph compared to Figure~\ref{fig:ekg}, a portion of which is illustrated in Figure~\ref{fig:tekg}. For instance, we have included $\mathsf{df}$ edges for snapshots. To avoid over-complicating the illustration, we have omitted drawing previous edges. In particular, we depict $\mathsf{corr}$ edges only for the snapshots of $\mathsf{a1}$, which caused the creation of $\mathsf{df}$ edges associated with the snapshots of $\mathsf{a1}$.

The tEKG in Figure~\ref{fig:tekg} is more feature rich than a pure EKG,
allowing analysts to monitor temporal aspects of entities. For instance, event~$\mathsf{e2}$ is connected to the $\mathsf{(a1,t1)}$ snapshot and not $\mathsf{(a1,t4)}$, which highlights that the student read the assignment when it had 2 points. Notice also that such connections between events and snapshots of entities at specific points in time are only implicitly present in OCEL~2.0. Making them explicit in a tEKG enables direct access to them for temporal analysis of event logs.

\section{Formalization}\label{sec:formalization}
This section defines our notion of a tEKG. We begin with a recap of definitions
	of~%
OCEL and EKG that we build
on and that are adapted from the related
publications~\cite{berti2023ocelspecification,fahland2022process, angles2017foundations}.

\subsection{Preliminaries}\label{ssec:preliminaries}

\begin{definition}\label{def:univ}\normalfont 
    $\univ{\Sigma}$ is a \textbf{universe} consisting of the following, pairwise  disjoint sets~\cite{berti2023ocelspecification}:
       
\begin{tabular}{p{60mm}l}
$\univ{eid}$ is the universe of event identifiers,
&
$\univ{\mathit{val}}$ is the universe of attribute values,
\\
$\univ{oid}$ is the universe of object identifiers,
&
$\univ{\mathit{time}}$ is the universe of timestamps,
\\
$\univ{\mathit{etype}}$ is the universe of event types,
&
$\univ{\mathit{qual}}$ is the universe of qualifiers, and
\\
$\univ{\mathit{otype}}$ is the universe of object types,
&
$\univ{\mathit{lbl}}$ is the universe of labels.
\\
$\univ{\mathit{att}}$ is the universe of attribute names,
&
\end{tabular}

\end{definition}

\begin{definition}\label{def:ocel}
\normalfont
An \textbf{Object-Centric Event Log~(OCEL)}~$L$ is a tuple~$(E,$  $O,\mathit{EA},$ $\mathit{OA},$ $\mathit{evtype},\mathit{evid},\mathit{time},\mathit{objtype},\mathit{objid},\mathit{eatype},\mathit{oatype},\mathit{eaval},
\mathit{oaval},\mathit{E2O},\mathit{O2O})$
	where~\cite{berti2023ocelspecification}:
 \begin{itemize}
    \item $E$ and $O$ are disjoint sets of events and of objects, respectively,
    \item $\mathit{EA} \!\subseteq\! \univ{\mathit{att}}$ and $\mathit{OA} \!\subseteq\! \univ{\mathit{att}}$ are sets of
    		attributes for events and objects, respectively,
    \item $\mathit{evtype}: E \rightarrow \univ{\mathit{etype}}$ is a function that assigns event types to events,
    \item $\mathit{evid}: E \rightarrow \univ{\mathit{eid}}$ is a function that assigns event id to events,
    \item $\mathit{time}: E \rightarrow \univ{\mathit{time}}$ is a function that assigns timestamps to events, 
    \item $\mathit{objtype}: O \rightarrow \univ{\mathit{otype}}$ is a function that assigns object types to objects, 
    \item $\mathit{objid}: O \rightarrow \univ{\mathit{oid}}$ is a function that assigns object id to objects,
    \item $\mathit{eatype}: \mathit{EA} \rightarrow \univ{\mathit{etype}}$ is a function that assigns event types to event attributes, 
    \item $\mathit{oatype}: \mathit{OA} \rightarrow \univ{\mathit{otype}}$ is a function that assigns object types to object attributes, 
    \item $\mathit{eaval}: (E \times \mathit{EA}) \not\rightarrow \univ{\mathit{val}}$ is a partial function that assigns values to (some) event attributes such that $evtype(e) = eatype(ea)$ for all $(e,ea)\in dom(eaval)$,

    \item $oaval: (O \times \mathit{OA} \times \univ{\mathit{time}}) \not\rightarrow \univ{\mathit{val}}$ assigns values to object attributes such that $objtype(o)=oatype(oa)$ for all $(o,oa,t)\in dom(oaval)$,

    \item $\mathit{E2O}\subseteq E \times \univ{\mathit{qual}} \times O$ are the qualified event-to-object relations, and

    \item $\mathit{O2O}\subseteq O \times \univ{\mathit{qual}} \times O$ are the qualified object-to-object relations.
    
 \end{itemize}
    
\end{definition}

While the $\mathit{oaval}$ function of OCEL assigns values to object attributes at particular points in time, the idea is that such a value remains valid until the next time point at which $\mathit{oaval}$ assigns a new value to the attribute. Yet, for all time points in between, the $\mathit{oaval}$ function is undefined for the corresponding attribute. To denote the \emph{current value} that an object $o\in O$ has for an attribute $\mathit{oa}\in\univ{\mathit{att}}$ at some arbitrary point in time $t\in\univ{\mathit{time}}$, the OCEL specification writes $\mathit{oaval}^t_{\mathit{oa}}(o)$, which we formalize~as~follows.

\begin{itemize}
	\item If there exists a timestamp $t' \in\univ{\mathit{time}}$ such that
	i)~$t' \leq t$,
	ii)~$(o,oa,t')\in dom(oaval)$, and
	iii)~there is no $t''\in\univ{\mathit{time}}$ such that $t^\prime<t''\leq t$ and $(o, \mathit{oa}, t'') \in dom(\mathit{oaval})$,
	then $\mathit{oaval}^t_{\mathit{oa}}(o)$ is $\mathit{oaval}(o, \mathit{oa}, t^\prime)$.

	\item If no such
	$t^\prime$ exists, then $\mathit{oaval}^t_{\mathit{oa}}(o)$ is $\perp$, where $\perp$ is a special value not in $\univ{\mathit{val}}$.
\end{itemize}
    
\begin{definition}\label{def:lpg}\normalfont A \textbf{Labeled Property Graphs~(LPG)}~$G$ is a tuple $(N, R, \gamma, \lambda, \rho )$ (adopted from ~\cite{fahland2022process, angles2017foundations}), where:
    \begin{itemize}
        \item $N$ and  $R$ are finite sets of nodes and of edges~(relationships), respectively,
        \item $\gamma: R \rightarrow N \times N$ is a function assigning pairs of source and target nodes to edges,
        \item $\lambda: (N \cup R)\rightarrow \univ{\mathit{lbl}}$ is a function assigning a label to every node or every edge,
        \item $\rho: (N\cup R) \times \univ{\mathit{att}} \nrightarrow \univ{\Sigma}\cup (\univ{\Sigma}\times\univ{\Sigma})\cup ((\univ{\Sigma}\times\univ{\Sigma})\times(\univ{\Sigma}\times\univ{\Sigma}))$ is a partial function assigning (potentially composite) values to attributes of node and edges.
    \end{itemize}
\end{definition}

Given an LPG $G=(N, R, \gamma, \lambda, \rho)$ and a label~$\ell \in \univ{\mathit{lbl}}$, we write $N^\ell$ to denote the subset of $N$ consisting of all the nodes with label $l$; i.e., $N^\ell = \{ n \in N \mid \lambda(n)=\ell\}$. Similarly, for edges: $R^\ell = \{ r \in R \mid \lambda(r)=\ell\}$.

We now introduce Event Knowledge Graphs (EKGs) as a special kind of LPGs that use a specific schema~$\mathcal{S}$,
	which we capture as a set of 3-tuples:
\begin{align*}
	\mathcal{S} \!=\! \big\{
	&
	(\mathsf{Log}, \mathsf{has}, \mathsf{Event}),
	(\mathsf{Event}, \mathsf{observed}, \mathsf{Class}),
	(\mathsf{Class}, \mathsf{dfc}, \mathsf{Class}),
	(\mathsf{Event}, \mathsf{df}, \mathsf{Event}),
	\\
	&
	(\mathsf{Event}, \mathsf{corr}, \mathsf{Entity}),
	(\mathsf{Entity}, \mathsf{rel}, \mathsf{Entity}),
	(\mathsf{Entity}, \mathsf{derived}, \mathsf{Entity})
\big\}.
\end{align*}

\noindent
Each such 3-tuple represents one of the types of edges in EKGs, where the second element of the 3-tuple provides the label of these edges, and the first and the third element captures the labels of corresponding source and target nodes, respectively. Formally, we say that an LPG~$G=(N, R, \gamma, \lambda, \rho)$ \emph{conforms to $\mathcal{S}$} if, for every edge~$r\in R$ with $\gamma(r)=(n,n')$, there exists $(s,l,t)\in S$ such that $n\in N^s$, $n' \in N^t$, and $r\in R^l$.

The $\rho$ function assigns values to an attribute of nodes and edges, so its range is defined to cover different scenarios that are informally explained in Figure~\ref{fig:tekg} such as a singular value (e.g. $a_1$), a tuple (e.g. $(a_1,t_1)$), a tuple of tuples (e.g. $((a_1,t_1),(a_1,t_4))$).

\begin{definition}\label{def:EKG}\normalfont
An \textbf{Event Knowledge Graph~(EKG)} is an LPG $(N, R, \gamma, \lambda, \rho)$ that conforms to the schema $\mathcal{S}$ and every node~$n \in N$ has the following properties
(as per~%
\cite{fahland2022process,esser2021multi}):
\begin{itemize}
	\item
	If $n \in N^{\mathsf{Event}}$, then $\rho(n,\mathsf{id})\in \univ{\mathit{eid}}$, $\rho(n,\mathsf{act}) \in \univ{\mathit{etype}}$, and $\rho(n,\mathsf{time}) \in \univ{\mathit{time}}$.

	\item
	If $n \in N^{\mathsf{Entity}}$, then $\rho(n,\mathsf{id}) \in \univ{\mathit{oid}}\cup(\univ{\mathit{oid}}\times\univ{\mathit{oid}})$ and $\rho(n,\mathsf{type}) \in \univ{\mathit{otype}}$.

\end{itemize}
\end{definition}

By Definition~\ref{def:EKG}, nodes with the label $\mathsf{Event}$ in an EKG have attributes $\mathsf{id}$, $\mathsf{act}$, and $\mathsf{time}$, with the value of an event identifier, an event type
and a timestamp, respectively.
Similarly, nodes with the label $\mathsf{Entity}$ have attributes $\mathsf{id}$ and $\mathsf{type}$, with the value of an object identifier and an object type, respectively. The $\mathsf{id}$ value can be a single identifier or a tuple thereof. Entities with an $\mathsf{id}$ value of the latter type are called \emph{reified entities}.

	In contrast to the original definition of EKGs~\cite{fahland2022process}, Definition~\ref{def:EKG} is more relaxed, as it does not enforce the existence of specific properties and edges. This flexibility allows our transformation algorithm to construct and add nodes, edges, and properties incrementally. The same approach is followed in the next definition.
	
\subsection{Temporal Event Knowledge Graphs}\label{ssec:tEKGs}
To define tEKGs that capture temporal objects, we extend the aforementioned schema~$\mathcal{S}$ by adding four more 3-tuples as follows:
\begin{align*}
\mathcal{S}' = \mathcal{S} \cup \{
&
(\mathsf{Event}, \mathsf{corr}, \mathsf{Snapshot}),
(\mathsf{Snapshot}, \mathsf{rel}, \mathsf{Snapshot}),
\\ &
(\mathsf{Entity}, \mathsf{snapshot},  \mathsf{Snapshot}),
(\mathsf{Snapshot}, \mathsf{derived}, \mathsf{Snapshot}) \} .
\end{align*}

\begin{definition}\label{def:tEKG}\normalfont A \textbf{temporal Event Knowledge Graph~(tEKG)}
is an LPG~%
	$(N, R,\gamma, \lambda, \rho)$
that conforms to the schema $\mathcal{S}'$ and every node~$n \in N$ has the properties as in an EKG (see Definition~\ref{def:EKG}) as well as the following property:
\begin{itemize}
	\item
	If $n \in N^{\mathsf{Snapshot}}$, then $ \rho(n,\mathsf{id}) \in (\univ{\mathit{oid}}\times\univ{\mathit{time}})\cup((\univ{\mathit{oid}}\times\univ{\mathit{time}})\times(\univ{\mathit{oid}}\times\univ{\mathit{time}}))$ and $\rho(n,\mathsf{type}) \in \univ{\mathit{otype}}$.
\end{itemize}
\end{definition}

By Definition~\ref{def:tEKG}, every node with the label $\mathsf{Snapshot}$ in a tEKG has attributes $\mathsf{id}$ and $\mathsf{type}$, with the value of a snapshot identifier and an object type, respectively. The snapshot identifier is a tuple of an object identifier and a time, or a tuple of such tuples. In the latter case, the corresponding snapshot is called a \emph{reified snapshot}.

\section{Transformation}\label{sec:transformation}

Given the notion of a tEKG, we now specify the transformation algorithm that converts logs from the OCEL~2.0 format into a corresponding tEKG representation. Algorithm~\ref{alg:map} defines the main part of the transformation, which is complemented by
\ConferenceVersion{%
	a procedure for creating the directly-follows edges (Algorithm~\ref{alg:processDF}). Additionally, the algorithm uses a procedure called \texttt{AddNode} for adding a node to the tEKG, and a procedure called \texttt{AddEdge} for adding an edge; due to space constraints, the detailed pseudo code for the latter two procedures is available only in the extended version of this paper~\cite{ExtendedVersion}.%
	\todo{Make sure that we have the extended version on arXiv and the bib entry refers to it.}%
}%
\ExtendedVersion{%
	procedures for creating the directly-follows edges (Algorithm~\ref{alg:processDF}) and for adding a node (Algorithm~\ref{alg:addnode}) and an edge (Algorithm~\ref{alg:addrelation}) to the tEKG. The input to the algorithm is an OCEL as per Definition~\ref{def:ocel} and the output is a tEKG as per Definition~\ref{def:tEKG}.
}

\begin{algorithm}[t!]
\caption{Converting an OCEL 2.0 log $L$ into a tEKG $G$.} \label{alg:map}
\KwData{
	$L = (E,O,\mathit{EA},\mathit{OA},\mathit{evtype},\mathit{evid},\mathit{time},\mathit{objtype},\mathit{objid},\mathit{eatype},\mathit{oatype},$
	\newline \hspace*{8mm}
	$\mathit{eaval},\mathit{oaval},\mathit{E2O},\mathit{O2O})$
}
\KwResult{$G=(N, R, \gamma, \lambda, \rho  )$}   

  \LinesNumberedHidden
    \LinesNumbered

    \SetKwBlock{Begin}{Begin}{}

        Create $G$ as an initially empty tEKG\label{line:initGraph}\;
        Let $\mathit{\wp} : E \cup O \cup \univ{\mathit{etype}} \cup 
        (O\times\univ{\mathit{time}}) \cup 
        (O\times O) \rightarrow N$ be an initially empty helper function that maps elements of $L$ to nodes created for them\label{line:function:wp}\;
        \tcp{add a node for the log}
        $N \leftarrow  N \cup \{\mathit{log}\}$, where $\mathit{log}$ is a new node that is not in $N$ (i.e., $\mathit{log} \notin N$)\label{line:log:start}\;
        Extend $\lambda$ such that $\lambda(\mathit{log})=\mathsf{log}$\label{line:log:end}\;
        \tcp{add a node for each event type}
        \ForEach{$c \in \univ{\mathit{etype}}$}{\label{line:class:start}
        $n\leftarrow$\texttt{{AddNode}{$(c, \mathsf{Class}, G, L)$}}\;
        Extend $\mathit{\wp}$ such that $\mathit{\wp}(c)=n$\;
        }\label{line:class:end}
        		\tcp{add a node for each event and connect it to both ...}
        \ForEach{$e \in E$}{\label{line:event:start}
        $n\leftarrow$\texttt{{AddNode}{$(e, \mathsf{Event}, G, L)$}}\;
        Extend $\mathit{\wp}$ such that $\mathit{\wp}(e)=n$\;
        \texttt{{AddEdge}{$(G,\mathit{log},n, \mathsf{has},\emptyset)$}} \tcp*{... the log node and the}
        \texttt{{AddEdge}{$(G,n,\mathit{\wp}(evtype(e)), \mathsf{observed},\emptyset)$}} \tcp*{node of its class}
        }\label{line:event:end}
        \tcp{add a node for each object}
        \ForEach{$o \in O$}{ \label{line:objects:start}
             $n\leftarrow$\texttt{{AddNode}{$(o, \mathsf{Entity}, G, L)$}}\;\label{line:entity}
            Extend $\mathit{\wp}$ such that $\mathit{\wp}(o) = n$\;\label{line:entity:fun}
            $\mathcal{O}_{\mathit{st}} \leftarrow \{ t\in\univ{\mathit{time}} | (o,\mathit{oa},t) \in dom(\mathit{oaval}) \text{ for some } oa\in \mathit{OA} \}$%
            	\;
				\label{line:object:changes}
            	\tcp{add a node for each object snapshot and connect ...}
            \ForEach{$t \in \mathcal{O}_{\mathit{st}}$ }{ \label{line:snapshot:start}
                $n^\prime\leftarrow$\texttt{{AddNode}{$((o,t), \mathsf{Snapshot}, G, L)$}}\;
                Extend $\mathit{\wp}$ such that $\mathit{\wp}((o,t)) = n^\prime$\;
                \texttt{{AddEdge}{$(G,n,n^\prime, \mathsf{snapshot},\emptyset)$}} \tcp*{... it to the object node}
            }\label{line:snapshot:end}
            \tcp{connect the object snapshots in their temporal order}
            \ForEach{$t_1,t_2 \in \mathcal{O}_{\mathit{st}}$}{ \label{line:update:start}
                \If{\normalfont $ t_1<t_2$ and there is no $t_3\in \mathcal{O}_{\mathit{st}}$ such that $t_1<t_3<t_2$}{
                    \texttt{{AddEdge}{$(G,\mathit{\wp}((o,t_1)),\mathit{\wp}((o,t_2)), \mathsf{rel},\mathsf{update})$}}\;
                }
            }\label{line:update:end}
        }  \label{line:objects:end}

        \tcp{connect objects and their snapshots using qualifiers}
        \ForEach{$(o_1,q,o_2) \in \mathit{O2O}$}{\label{line:o2o:start}
             \texttt{{AddEdge}{$(G, \mathit{\wp}(o_1), \mathit{\wp}(o_2), \mathsf{rel}, q)$}}\;\label{line:entity:rel}

            \ForEach{\normalfont$r\in R^\mathsf{snapshot}$ with $\gamma(r)=(\mathit{\wp}(o_1),\mathit{os}_1)$}{
                $\mathcal{R}_{\mathit{st}} \!\leftarrow\! \{ r'\! \!\in\! R^\mathsf{snapshot}\ |\  \rho(\mathit{os}_2,\mathsf{time}) \!\leq\! \rho(\mathit{os}_1,\mathsf{time}) $ with $\gamma(r') \!=\! (\wp(o_2),\mathit{os}_2) \}$\;\label{line:snapshot:set}

                \ForEach{\normalfont$r'\! \in \mathcal{R}_{\mathit{st}}$ for which there is no $r''\! \!\in\! \mathcal{R}_{\mathit{st}}$ with $\gamma(r'') \!=\!(\mathit{\wp}(o_2),\mathit{os}_2')$ such that $\rho(\mathit{os}_2,\mathsf{time}) \!<\! \rho(\mathit{os}_2^\prime,\mathsf{time})$}{ 
                \label{line:snapshot:foreach}
                    \tcp{connect existing snapshots at a time ... }
                    \texttt{{AddEdge}{$(G,\mathit{os}_1,\mathit{os}_2, \mathsf{rel}, q)$}}\;\label{line:snapshot:rels} 
                }
            }
        }
\end{algorithm}

\begin{algorithm}[t]
    \setcounter{AlgoLine}{31}

    \SetKwBlock{Begin}{}{end}
        \tcp{add nodes for reified entities and reified snapshots}
        \ForEach{\normalfont$r \in R^\mathsf{rel} $ with $ \gamma(r)=(\mathit{\wp}(o_1),\mathit{\wp}(o_2))$}{\label{line:reified:loop}
                $n\leftarrow$\texttt{{AddNode}{$((o_1,o_2), \mathit{label}, G, L)$}}, where $\mathit{label} = \lambda(\mathit{\wp}(o_1))$\;
                Extend $\mathit{\wp}$ such that $\mathit{\wp}((o_1,o_2))=n$\;
                \texttt{{AddEdge}{$(G,n,\mathit{\wp}(o_1), \mathsf{derived},\emptyset)$}}\;
                \texttt{{AddEdge}{$(G,n,\mathit{\wp}(o_2), \mathsf{derived},\emptyset)$}} \;
        }\label{line:reified:loop:end}

        \ForEach{$(e,q,o) \in \mathit{E2O}$}{\label{line:e2o:start}
            \tcp{connect event nodes to corresponding entity nodes}
            \texttt{{AddEdge}{$(G,\mathit{\wp}(e),\mathit{\wp}(o), \mathsf{corr}, q)$}}\;\label{line:e2o:edge}
            \ForEach{\normalfont$r\in R^\mathsf{derived}$ with $ \gamma(r)=(o'\!,o)$}{ 
                            \texttt{{AddEdge}{$(G,\mathit{\wp}(e), o'\!, \mathsf{corr}, q)$}}\; \label{line:e2o:edge:refified}
                    }

            \tcp{connect event nodes to corresponding snapshot nodes}
            $\mathcal{R}_{\mathit{st}}\leftarrow \{ r\in R^\mathsf{snapshot}\ |\  \rho(\mathit{os}_1,\mathsf{time})~\leq~\rho(\mathit{\wp}(e),\mathsf{time})$ \text{with} $\gamma(r)=(\wp(o),\mathit{os}_1)\}$\;\label{line:e2o:snapshots}
            \ForEach{\normalfont$r \in \mathcal{R}_{\mathit{st}}$ with $\gamma(r)=(\wp(o),\mathit{os}_1)$}{%
                \If{\normalfont there is no $
                        r'\! \in \mathcal{R}_{\mathit{st}}$ with $
                        \gamma(r')=(\mathit{\wp}(o),\mathit{os}_2) 
                        $ such that $
                        \rho(\mathit{os}_1,\mathsf{time}) < \rho(\mathit{os}_2,\mathsf{time})$}{
                    \texttt{{AddEdge}{$(G,\mathit{\wp}(e),\mathit{os}_1, \mathsf{corr}, q)$}}\;\label{line:e2o:snapshots:edge}
                    \ForEach{\normalfont$r''\! \in R^\mathsf{derived}$ with $ \gamma(r'')=(\mathit{os}_3,\mathit{os}_1)$ }{
                            \texttt{{AddEdge}{$(G,\mathit{\wp}(e),\mathit{os}_3, \mathsf{corr}, q)$}}\;\label{line:e2o:snapshots:edge:reified}
                    }
                }
            }
        }\label{line:e2o:end}

        $G\leftarrow$\texttt{{AddDFs}{$(G)$}} \tcp*{add directly-follows edges} \label{line:adding:dfs}
        \KwRet{G}\;
\end{algorithm}

After initializing the tEKG to be populated~(line~\ref{line:initGraph} in Algorithm~\ref{alg:map}), the algorithm initializes a data structure---captured by function $\mathit{\wp}$---for tracking the nodes in the tEKG that have been created based on specific elements of the log (line~\ref{line:function:wp}). Next, a node for the log itself is added to the tEKG (lines~\ref{line:log:start}--\ref{line:log:end}). The next step involves adding a node with label~$\mathsf{Class}$ to the tEKG for each event type in the log (lines~\ref{line:class:start}--\ref{line:class:end}).
After that, for each event in the log, a node with label~$\mathsf{Event}$ is added and connected to the previously-created nodes for both the log and the corresponding event type (lines~\ref{line:event:start}--\ref{line:event:end}).

The algorithm then iterates over all objects in the log%
~(lines~\ref{line:objects:start}--\ref{line:objects:end}).
	For each object, it adds a node with label~$\mathsf{Entity}$ (line~\ref{line:entity}). 
Next, it identifies all timestamps at which the value of an attribute of the object has changed (line~\ref{line:object:changes}).
The algorithm then iterates over these timestamps, adding a node with label~$\mathsf{Snapshot}$ to the tEKG and linking it to the corresponding entity (lines~\ref{line:snapshot:start}--\ref{line:snapshot:end}).
Snapshots are added if the object has a value over time. If an object has no attribute and value in OCEL (i.e., it is not initiated or related to an event), the algorithm
	creates no
snapshot for it.
Finally, the algorithm creates edges between such snapshot nodes to represent the updates occurring over time~(lines~\ref{line:update:start}--\ref{line:update:end}).

The next step involves adding a $\mathsf{rel}$-labeled edge for every object-to-object relationship between any two objects $o_1$ and $o_2$~(line~\ref{line:entity:rel}).
After that, the algorithm iterates over all snapshot edges created for the node corresponding to $o_1$ in tEKG. It then collects all snapshots of the node corresponding to $o_2$ in tEKG that occurred before the snapshot of $o_1$, and puts them into a set named $\mathcal{R}_{\mathit{st}}$ (line~\ref{line:snapshot:set}). This set is used to filter the last valid snapshot for $o_2$ at the time of $o_1$, to which we can link the snapshot~(line~\ref{line:snapshot:rels}). 
In our running example~(see Figure~\ref{fig:tekg}), the snapshot $(c1,t1)$ could be linked to $(a1,t1)$, which is a snapshot of the related object with a timestamp that is less than or equal to $(c1,t1)$. 

\begin{algorithm}[t!]
\caption{Extending a given tEKG with directly-follows edges.}\label{alg:processDF}
  \SetKwProg{Fn}{Function}{:}{}
  \SetKwFunction{FAddDFs}{AddDFs}
  \Fn{\FAddDFs} {
      \KwData{$G=(N,R,\gamma,\lambda,\rho)$}
      \KwResult{$G$, extended with directly-follows edges}
        \tcp{add directly-follows edges between event nodes}
         \ForEach{\normalfont%
         	$r_1,r_2 \!\in\! R^\mathsf{corr}$ with $\gamma(r_1) \!=\! (e_1,o)$ and $\gamma(r_2) \!=\! (e_2,o)$ such that $e_1 \!\neq\! e_2$}
         {%
             \If{\normalfont there is no $
                            r_3\in R^\mathsf{corr}$ with $
                            \gamma(r_3)=(e_3,o)
                            $ such that $ e_1\neq e_2\neq e_3 $ and $ 
                            \rho(e_1,\mathsf{time})~<~\rho(e_3,\mathsf{time})~<~\rho(e_2,\mathsf{time})$}{
                $r \leftarrow$ \texttt{{AddEdge}{$(G,e_1,e_2, \mathsf{df}, \emptyset)$}}\;\label{line:df:edge}
                Extend $\rho$ such that $\rho\big(r,\mathsf{type}\big)=\rho(o,\mathsf{type})$ and $\rho\big(r,\mathsf{ent}\big)=\rho(o,\mathsf{id})$\;\label{line:df:edge:typeandent}
                }
            } 
        \tcp{identify directly-follows edges providing new information}
        $I \leftarrow \emptyset$\; \label{line:df:addedinfo:start}
        \ForEach{$\mathit{label}\in\{\mathsf{Entity},\mathsf{Snapshot}\}$}  {
            \ForEach{\normalfont$r \in R^\mathsf{df}$ \text{and} $o\in N^{\mathit{label}}$ \text{such that} $\rho(r,\mathsf{ent})=\rho(o,\mathsf{id})$}
            {
                \If{\normalfont there is no $r'\! \in R^\mathsf{df}$ and $o'\! \in N^{\mathit{label}}$ and $ r''\! \in R^\mathsf{derived}$ such that
                $\rho(r'\!,\mathsf{ent})=\rho(o'\!,\mathsf{id})$ and $ \gamma(r)=\gamma(r')$ and $\gamma(r'')=(o,o')$}
                {
                    $I \leftarrow I \cup \{r\}$\;
                }
            }
        }\label{line:df:addedinfo:end}
        \tcp{remove directly-follows edges not providing new information}
        
        \ForEach{\normalfont%
        		$r_1,r_2 \!\in\! R^\mathsf{df}$ such that $r_1 \!\neq\! r_2$ and $\gamma(r_1) \!=\! \gamma(r_2)$, with $\gamma(r_1) \!=\! (e_1,e_2)$}
        {%
            \If{$r_1 \not\in I$}
            {
                \If{\normalfont there are no
                	$r_3, r_4\in R^\mathsf{df}$ with $\gamma(r_3)=(e'\!,e_1)$ and $\gamma(r_4)=(e_2,e'')$ such that $\rho(r_1,\mathsf{ent})=\rho(r_2,\mathsf{ent})=\rho(r_3,\mathsf{ent})$ and $r_3\in I$ and $r_4\in I$}
                {
                    $R \leftarrow R \char`\\  \{r_1\}$\;
                    $dom(\gamma) \leftarrow dom(\gamma) \char`\\  \{r_1\}$\;
                    $dom(\lambda) \leftarrow dom(\lambda) \char`\\  \{r_1\}$\;
                    \ForEach{$\mathit{att} \in \univ{\mathit{att}}$}
                    {
                        \If{$(r_1, \mathit{att})\in dom(\rho)$}
                        {
                            $dom(\rho) \leftarrow dom(\rho) \char`\\  \{(r_1, \mathit{att})\}$\;
                        }
                    }
                }                      
            }
        }
        \KwRet{$G$}\;
    }
\end{algorithm}

The next step focuses on reified entities and reified snapshots. Here, the algorithm iterates over all
	$\mathsf{rel}$-labeled edges in the tEKG
and adds a node for each of them, as well as an edge from this node to the start and end node of each of these edges~(lines~\ref{line:reified:loop}--\ref{line:reified:loop:end}).

	Next,
the algorithm iterates over all
	event-to-object relationships
in the log (lines~\ref{line:e2o:start}--\ref{line:e2o:end})%
	, performing the following operations for each of them:
\textit{First}, it adds an edge between the corresponding event and entity
(line~\ref{line:e2o:edge}).
\textit{Second}, it iterates over all reified entities derived from the corresponding entity and adds an edge from the corresponding event to each of them to tEKG aligned with design choice made in~\cite{esser2021multi} (line~\ref{line:e2o:edge:refified}).
\textit{Third}, it retrieves a set of snapshots for the corresponding object that existed at the time of the event (line~\ref{line:e2o:snapshots}).
\textit{Fourth}, it connects the corresponding event to the last valid snapshot (line~\ref{line:e2o:snapshots:edge}), as well as connecting the event to all derived snapshots of the given snapshot aligning with the same design choice for reified entities made in~\cite{esser2021multi}~(line~\ref{line:e2o:snapshots:edge:reified}).

In the end, Algorithm~\ref{alg:processDF} is called
(line~\ref{line:adding:dfs}).
This algorithm receives
	the current
tEKG as input, adds relevant directly-follows edges to it, and returns the updated graph as output. The algorithm consists of three \textit{phases}: adding all directly-follows edges, identifying edges that add new information%
, and removing the ones that do not.

	More specifically, in the \textit{first phase}, the algorithm iterates over any two $\mathsf{corr}$ edges that are targeting the same entity from two events. If there are no other events occurring in between that have a $\mathsf{corr}$ edge to the same entity, the algorithm adds an edge with label~$\mathsf{df}$ between those two events~(line~\ref{line:df:edge}). It also sets the value of the $\mathsf{ent}$ and $\mathsf{type}$ attributes of the added edge to the value of $\mathsf{id}$ and $\mathsf{type}$ of the entity (line~\ref{line:df:edge:typeandent}).

The \textit{second phase} identifies all $\mathsf{df}$-labeled edges
that provide new information (lines \ref{line:df:addedinfo:start}--\ref{line:df:addedinfo:end}). To this end, the algorithm applies the same rule as defined by Fahland~\cite{fahland2022process}, stating that not all $\mathsf{df}$ edges created for derived entities provide additional information. Specifically, if there is a derived node $o$ related to $o^\prime$ for which there exist $\mathsf{df}$ edges between two events, the $\mathsf{df}$ edge created for the derived entity $\mathsf{o}$ does not add new information.

In the \textit{last phase}, the algorithm removes $\mathsf{df}$ edges that do not add new information with the condition that there shall not be any similar $\mathsf{df}$ edges both before and after them that are among the added information $\mathsf{df}$.

\ExtendedVersion{
	Algorithm~\ref{alg:addnode}
		is
	a helper algorithm responsible for adding a node to the tEKG. This algorithm takes input information from OCEL to set the properties of the newly-created node based on its types.
	Algorithm~\ref{alg:addrelation} adds an edge between two nodes to the~tEKG.
}

\section{Concluding Remarks}\label{sec:conclusion}

This paper introduces and formalizes temporal Event Knowledge Graphs (tEKGs), which are designed to record object-centric event
	data and to facilitate
the tracking of changes in entity attributes over time. For instance, consider the price of item, which can fluctuate; tEKGs allow for the analysis of events with respect to the accurate price at any given time, before or after any changes. This capability is crucial for conducting effective data-driven analyses in real-world scenarios.
Moreover, the paper presents
	an
algorithm
	to transform
Object-Centric Event Logs (OCEL) 2.0 into tEKGs. 

As a future direction, we aim to provide a complete formal definition of temporal event knowledge graphs by eliciting requirements for object-centric event data based on different case studies. Investigating the practical applications of tEKGs could provide deeper insights into business processes and decision making.

\subsubsection*{Acknowledgements.}
    Khayatbashi's and Hartig's
contributions to this work
were funded by Vetenskapsrådet (the Swedish Research Council, project
reg. no. 2019-05655).

\bibliographystyle{abbrv} 
\bibliography{main} 

\begin{thebibliography}{10}

\bibitem{angles2017foundations}
R.~Angles, M.~Arenas, P.~Barcel{\'o}, A.~Hogan, J.~Reutter, and D.~Vrgo{\v{c}}.
\newblock {Foundations of Modern Query Languages for Graph Databases}.
\newblock {\em ACM Computing Surveys}, 50(5), 2017.

\bibitem{berti2023ocelspecification}
A.~Berti, I.~Koren, J.~N. Adams, G.~Park, B.~Knopp, N.~Graves, M.~Rafiei,
  L.~Li{\ss}, L.~T.~G. Unterberg, Y.~Zhang, C.~Schwanen, M.~Pegoraro, and
  W.~{van der Aalst}.
\newblock {OCEL} ({O}bject-{C}entric {E}vent {L}og) 2.0 specification.
\newblock
  {\footnotesize\url{https://www.ocel-standard.org/2.0/ocel20_specification.pdf}},
  2023.

\bibitem{esser2019storing}
S.~Esser and D.~Fahland.
\newblock {Storing and Querying Multi-Dimensional Process Event Logs using
  Graph Databases}.
\newblock In {\em Business Process Management Workshops: BPM 2019 International
  Workshops, Vienna, Austria, September 1--6, 2019, Revised Selected Papers
  17}, 2019.

\bibitem{esser2021multi}
S.~Esser and D.~Fahland.
\newblock Multi-dimensional event data in graph databases.
\newblock {\em Journal on Data Semantics}, 10(1-2):109--141, 2021.

\bibitem{fahland2022process}
D.~Fahland.
\newblock Process mining over multiple behavioral dimensions with event
  knowledge graphs.
\newblock In {\em Process Mining Handbook}, pages 274--319. Springer, 2022.

\bibitem{ghahfarokhi2021ocel}
A.~F. Ghahfarokhi, G.~Park, A.~Berti, and W.~M. van~der Aalst.
\newblock {OCEL: A Standard for Object-Centric Event Logs}.
\newblock In {\em New Trends in Database and Information Systems: ADBIS 2021
  Short Papers, Doctoral Consortium and Workshops}, 2021.

\bibitem{gherissi2023object}
W.~Gherissi, J.~El~Haddad, and D.~Grigori.
\newblock Object-centric predictive process monitoring.
\newblock In {\em International Conference on Service-Oriented Computing},
  pages 27--39. Springer, 2022.

\bibitem{jalali2022object}
A.~Jalali.
\newblock Object type clustering using markov directly-follow multigraph in
  object-centric process mining.
\newblock {\em IEEE Access}, 10:126569--126579, 2022.

\bibitem{jalali2013multi}
A.~Jalali and P.~Johannesson.
\newblock Multi-perspective business process monitoring.
\newblock In {\em Int.\ Workshop on Business Process Modeling, Development and
  Support}, 2013.

\bibitem{khayatbashi2023transforming}
S.~Khayatbashi, O.~Hartig, and A.~Jalali.
\newblock Transforming event knowledge graph to object-centric event logs: A
  comparative study for multi-dimensional process analysis.
\newblock In {\em International Conference on Conceptual Modeling}, pages
  220--238. Springer, 2023.

\bibitem{kimball2011data}
R.~Kimball and M.~Ross.
\newblock {\em The data warehouse toolkit: the complete guide to dimensional
  modeling}.
\newblock John Wiley \& Sons, 2011.

\bibitem{klijn2022aggregating}
E.~L. Klijn, F.~Mannhardt, and D.~Fahland.
\newblock Aggregating event knowledge graphs for task analysis.
\newblock In {\em International Conference on Process Mining}, pages 493--505.
  Springer, 2022.

\bibitem{swevels2023inferring}
A.~Swevels, R.~Dijkman, and D.~Fahland.
\newblock Inferring missing entity identifiers from context using event
  knowledge graphs.
\newblock In {\em Int.\ Conference on Business Process Management}, 2023.

\bibitem{van2023object}
W.~{van der Aalst}.
\newblock Object-centric process mining: Unraveling the fabric of real
  processes.
\newblock {\em Mathematics}, 11(12):2691, 2023.

\bibitem{van2020discovering}
W.~{van der Aalst} and A.~Berti.
\newblock Discovering object-centric petri nets.
\newblock {\em Fundamenta informaticae}, 175(1-4):1--40, 2020.

\end{thebibliography}

\ExtendedVersion{
	\appendix
	\newpage
	\section{Algorithms for Creating Nodes and Edges in the tEKG}
	\begin{algorithm}[b!]
\caption{Adding a node with relevant properties to a tEKG}\label{alg:addnode}
  \SetKwFunction{FAddNode}{AddNode}
  \SetKwProg{Fn}{Function}{:}{}
  \Fn{\FAddNode} {
    \KwData{$ 
        x\in E \cup O \cup \univ{\mathit{etype}} \cup 
        (O\times\univ{\mathit{time}}) \cup 
        (O\times O) \cup ((O\times\univ{\mathit{time}})\times(O\times\univ{\mathit{time}})) $; \newline
        $ l\in\univ{\mathit{lbl}}$;  \newline
        $G=(N, R, \gamma, \lambda, \rho  )$; \newline
        $L=(E,O,\mathit{EA},\mathit{OA},\mathit{evtype},\mathit{evid},\mathit{time},\mathit{objtype},\mathit{objid},\mathit{eatype},\mathit{oatype},$ \newline \hspace*{7mm} $\mathit{eaval},\mathit{oaval},\mathit{E2O},\mathit{O2O})$}
    \KwResult{the newly-create node}
        $N \leftarrow  N \cup \{n\}$, where $n$ is a new node that is not in $N$ (i.e., $n \notin N$)\;
        Extend $\lambda$ such that $\lambda(n)=l$\;
        \If{ $x\in E$}{
            Extend $\rho$ such that $\rho(n,\mathsf{id})=\mathit{evid}(x)$, 
            \, $\rho(n,\mathsf{act})=\mathit{evtype}(x)$,
            \newline
            $\rho(n,\mathsf{time})=\mathit{time}(x)$, and $\rho(n,\mathit{att})=\mathit{eaval}(x,\mathit{att})$ for all
            \newline
            $\mathit{att} \in \mathit{EA}$ for which $(x,\mathit{att}) \in dom(\mathit{eaval})$\;
        }
        \If{ $x\in O$}{
            Extend $\rho$ such that $\rho(n,\mathsf{id})=\mathit{objid}(x)$ and $\rho(n,\mathsf{type})=\mathit{objtype}(x)$\;
        }
        \If{ $x\in \univ{\mathit{etype}}$}{
            Extend $\rho$ such that $\rho(n,\mathsf{type})= x$ and $\rho(n,\mathsf{id})= x$\;
        }
        \If{ $x = (x^\prime,t)\in O\times\univ{\mathit{time}}$}{
            Extend $\rho$ such that $\rho(n,\mathsf{id})=(\mathit{objid}(x^\prime),t)$, \, $\rho(n,\mathsf{type}) = \mathit{objtype}(x^\prime)$,
            \newline
            $\rho(n,\mathsf{time})=t$, and $\rho(n,\mathit{oa})=\mathit{oaval}^t_{\mathit{oa}}(o,\mathit{oa})$ for all $oa \in \mathit{OA}$ for which $\mathit{oatype}(\mathit{oa})=\mathit{objtype}(o)$\;     
        }
        \If{ $x = (x_1,x_2) \in O\times O$}{
            Extend $\rho$ such that $\rho(n,\mathsf{id})=(\mathit{objid}(x_1),\mathit{objid}(x_2))$ and 
            $\rho(n,\mathsf{type})=(\mathit{objtype}(x_1),\mathit{objtype}(x_2))$\;
        }
        \If{ $x = ((x_1,t_1),(x_2,t_2)) \in ((O\times\univ{\mathit{time}})\times(O\times\univ{\mathit{time}})) $}{
            Extend $\rho$ such that $\rho(n,\mathsf{id})=((\mathit{objid}(x_1),t_1),(\mathit{objid}(x_2),t_2))$ and 
            $\rho(n,\mathsf{type})=(\mathit{objtype}(x_1),\mathit{objtype}(x_2))$\;
        }
        \KwRet{n}\;
    }
\end{algorithm}

	\vspace*{-5mm}
	\begin{algorithm}[b!]
\caption{Adding an edge to a tEKG}\label{alg:addrelation}
  \SetKwProg{Fn}{Function}{:}{}
  \SetKwFunction{FAddEdge}{AddEdge}
  \Fn{\FAddEdge} {
      \KwData{$G=(N,R,\gamma,\lambda,\rho), \ 
            n_1\in N, \ 
            n_2\in N,\ 
            l\in\univ{\mathit{lbl}}, \ 
            q\in\univ{qual}\cup\{\emptyset\}$}
      \KwResult{the newly-created edge}
        $R \leftarrow R \cup \{r\}$, where $r$ is a new edge that is not in $R$ (i.e., $r \notin R$) \; 
        Extend $\gamma$ such that $\gamma(r)=(n_1,n_2)$\;
        Extend $\lambda$ such that $\lambda(r)=l$\;
        Extend $\rho$ such that $\rho(r,\mathsf{id})=(\rho(n_1,\mathsf{id}),\rho(n_2,\mathsf{id}))$\;
        \If{ $q\neq\emptyset$}{
            Extend $\rho$ such that $\rho\big(r, \mathsf{qual}\big)=q$\;
        }
        \KwRet{r}\;
    }
\end{algorithm}

}

\end{document}